\documentclass{PoS} \usepackage{epsfig}

\title{Critical Exponents for U(1) Lattice Gauge Theory at Finite 
       Temperature}

\ShortTitle{Critical Exponents for U(1) Lattice Gauge Theory at 
            Finite Temperature}

\author{\speaker{Bernd A. Berg} and Alexei Bazavov\\
        Florida State University, Department of Physics,
        Tallahassee, FL 32306-4350, USA\\
        and\\
        Florida State University, School of Computational Science,
        Tallahassee, FL 32306-4120, USA\\
        E-mail: \email{berg@scs.fsu.edu},
        \email{bazavov@scs.fsu.edu}}

\abstract{
For compact U(1) lattice gauge theory (LGT) we have performed a 
finite size scaling analysis 
on $N_{\tau} N_s^3$ lattices for $N_{\tau}$ fixed and $N_s\to\infty$,
approaching the phase transition from the confined phase. For 
$N_{\tau}=4$, 5 and 6 our data contradict the expected scenario that 
this transition is either first order or in the universality class 
of the 3d XY model. If there are no conceptional flaws in applying the 
argument that the Gaussian fixed point in 3d is unstable to our 
systems, estimates of the critical exponents $\alpha/\nu$, 
$\gamma/\nu$, $(1-\beta)/\nu$ and $2-\eta$ indicate the existence 
of a new, non-trivial renormalization group fixed point for second 
order phase transitions in 3d. Such a fixed point would be of 
importance for renormalization group theory and statistical physics.}

\FullConference{XXIVth International Symposium on Lattice Field Theory\\
                July 23-28, 2006\\
                Tucson, Arizona, USA}

\begin{document}

\section{Model and Simulation Statistics}

This paper summarizes and extends work of Ref.~\cite{BeBa06} about pure 
U(1) LGT with the Wilson action. Multicanonical simulations \cite{BeNe91} 
on 
\begin{equation} \label{lattices}
  N_{\tau}\,N_s^3~~~{\rm lattices,}~~~N=N_{\tau}\,N_s^3
\end{equation}
are performed with $N_{\tau} = {\rm fixed},\ N_s\to\infty$ or 
$N_{\tau}=N_s$. The multicanonical parameters were determined
using a modification of the Wang-Landau (WL) recursion \cite{WaLa01}.
A speed up by a factor of about three was achieved by implementing
the biased Metropolis-Heatbath algorithm \cite{BaBe05} for the updating 
instead of relying on the usual Metropolis procedure. Additional 
overrelaxation \cite{Ad81} sweeps are presently used for simulations 
on larger lattices. 

Our statistics is summarized in table~\ref{tab_stat}. The lattice sizes 
are collected in the first and second column. The third column contains 
the number of sweeps spent on the WL recursion for the multicanonical 
parameters. Typically the parameters are frozen after reaching 
$\epsilon = e^{1/20}$. Exceptions are the simulations for which 
the values are marked by $*$ for which $\epsilon = e^{1/22}$ was 
used (technical details will be published elsewhere). Column four 
lists our production statistics from simulations with fixed 
multicanonical weights. Error bars as shown in figures will be 
calculated using jackknife bins (e.g., chapter~2.7 of \cite{BBook}) 
with their number given by the first 
value in column four, while the second value was also used for the 
number of equilibrium sweeps (without measurements) performed after 
the recursion. Columns five and six give the $\beta$ values between 
which our Markov process cycled. Adapting the definition of chapter~5.1 
of \cite{BBook} one cycle takes the process from the configuration 
space region at $\beta_{\min}$ to $\beta_{\max}$ and back. Each run 
was repeated once more, where after the first run the multicanonical 
parameters were estimated from its statistics. Columns seven and eight 
give the number of cycling events recorded during these runs. We are 
still producing on lattices with larger $N_s$ values than those listed 
in table~\ref{tab_stat} and are creating entire new series of $N_s$ 
values for $N_{\tau}=2,$ 3, 8 and~10.
We perform finite size scaling (FSS) calculations for the critical 
exponents of U(1) LGT. For a review of FSS methods and scaling relations 
see~\cite{PeVi02}. The observables, which we calculated, and their FSS 
behavior are introduced in the following.

\begin{table}[hp]
\caption{ Statistics.  \label{tab_stat}} \medskip
\centering
\begin{tabular}{|c|c|c|c|c|c|c|c|} \hline 
$N_{\tau}$ & $N_s$ & WL & Production & 
$\beta_{\min}$ & $\beta_{\max}$ & cycles 1 & cycles 2 \\ \hline
 4&  4&11$\,$592& $32\times  20$\,$000$& 0.8  & 1.2  & 527 & 594\\ \hline
 4&  5&14$\,$234& $32\times  12$\,$000$& 0.8  & 1.2  & 146 & 172\\ \hline
 4&  6&19$\,$546& $32\times  32$\,$000$& 0.9  & 1.1  & 258 & 364\\ \hline
 4&  8&29$\,$935& $32\times  32$\,$000$& 0.95 & 1.05 & 229 & 217\\ \hline
 4& 10&25$\,$499& $32\times  64$\,$000$& 0.97 & 1.03 & 175 & 317\\ \hline
 4& 12&47$\,$379&$32\times  112$\,$000$& 0.98 & 1.03 & 338 & 360\\ \hline
 4& 14&44$\,$879&$32\times  112$\,$000$& 0.99 & 1.02 & 329 & 322\\ \hline
 4& 16&54$\,$623&$32\times  128$\,$000$& 0.99 & 1.02 &  19 & 219\\ \hline
 4& 18&58$\,$107&$32\times  150$\,$000$& 0.994& 1.014&  93 & 259\\ \hline
 4& 22&73$\,$874*&$64\times  100$\,$000$&1.000& 1.008& 335 & 356\\ \hline
 5&  5&18$\,$201& $32\times  12$\,$000$& 0.8  & 1.2  & 114 & 122\\ \hline
 5&  6&20$\,$111& $32\times  36$\,$000$& 0.9  & 1.1  & 294 & 308\\ \hline
 5&  8&31$\,$380& $32\times  40$\,$000$& 0.95 & 1.05 &  35 & 191\\ \hline
 5& 10&47$\,$745& $32\times  72$\,$000$& 0.97 & 1.03 & 144 & 231\\ \hline
 5& 12&37$\,$035& $32\times 112$\,$000$& 0.99 & 1.02 & 280 & 326\\ \hline
 5& 14&49$\,$039& $32\times 112$\,$000$& 1.0  & 1.02 & 192 & 277\\ \hline
 5& 16&43$\,$671& $32\times 160$\,$000$& 1.0  & 1.02 & 226 & 257\\ \hline
 5& 18&56$\,$982&$32\times  180$\,$000$& 1.0  & 1.014& 138 & 241\\ \hline
 6&  6&28$\,$490& $32\times  40$\,$000$& 0.9  & 1.1  & 312 & 281\\ \hline
 6&  8&44$\,$024& $32\times  40$\,$000$& 0.96 & 1.04 & 173 & 175\\ \hline
 6& 10&51$\,$391& $32\times  72$\,$000$& 0.97 & 1.04 & 139 & 170\\ \hline
 6& 12&41$\,$179& $32\times 128$\,$000$& 0.995& 1.02 & 226 & 283\\ \hline
 6& 14&50$\,$670& $32\times 128$\,$000$& 1.0  & 1.02 &  89 & 220\\ \hline
 6& 16&56$\,$287& $32\times 160$\,$000$& 1.0  & 1.02 & 149 & 189\\ \hline
 6& 18&68$\,$610& $32\times  180$\,$000$&1.005& 1.015& 123 & 200\\ \hline
 8&  8&46$\,$094& $32\times  40$\,$000$& 0.97 & 1.03 & 111 & 159\\ \hline
10& 10&48$\,$419& $32\times  96$\,$000$& 0.98 & 1.03 & 103 & 133\\ \hline
12& 12&70$\,$340& $32\times 112$\,$000$& 0.99 & 1.03 &  75 &  82\\ \hline
14&14&112$\,$897& $32\times 128$\,$000$& 1.0  & 1.02 &  57 &  51\\ \hline
16& 16&87$\,$219& $32\times 160$\,$000$& 1.007& 1.015&  12 &  73\\ \hline
16&16&191$\,$635*&$32\times 160$\,$000$& 1.007& 1.015&  48 &  74\\ \hline
\end{tabular} \end{table} 

For the Specific heat one has
\begin{equation} \label{Cbeta}
  C(\beta)\ =\ \frac{1}{6N} \left[\langle S^2\rangle -
  \langle S\rangle^2 \right]_{\max} \sim N_s^{\alpha/\nu}\ .
\end{equation} 
Polyakov loops $P_{\vec{x}}$ are the products of U(1) gauge matrices 
along straight lines in the $N_{\tau}$ direction. The lattice average 
of Polyakov loops is defined by 
\begin{equation} \label{PLoop}
  P\ =\ \sum_{\vec{x}} P_{\vec{x}}\ .
\end{equation} 
The maxima of the susceptibility of the absolute value $|P|$ (called
Polyakov loop susceptibility henceforth) scale like
\begin{equation} \label{chi}
  \chi_{\max}\ =\ \frac{1}{N_s^3} \left[ \langle|P|^2\rangle -
  \langle|P|\rangle^2\ \right]_{\max} \sim\ N_s^{\gamma/\nu}\,,
\end{equation} 
while for the maxima of the derivatives the scaling behavior
\begin{equation} \label{chi_beta}
  \chi^{\beta}_{\max}\ =\ \frac{1}{N_s^3} \left. \frac{d~}{d\beta}\,
  \langle|P|\rangle\ \right|_{\max} \sim\ N_s^{(1-\beta)/\nu}\,.
\end{equation} 
holds.  Maxima of structure factors (see, e.g., Ref.~\cite{Stanley})
\begin{equation} \label{sf}
  F(\vec{k})=\frac{a}{N_s^3} \left\langle\left|\sum_{\vec{r}}
  P(\vec{r})\, \exp(i\vec{k}\vec{r}) \right|^2\right\rangle\ ,
  \qquad (a=1~~{\rm lattice\ constant})\ ,
\end{equation} 
scale like
\begin{equation} \label{sffss}
  F(\vec{k})_{\max}\ \sim\ N_s^{2-\eta}\ =\ N_s^{\gamma/\nu}\,.
\end{equation} 

\section{Estimates of Critical Exponents}

\begin{figure}[-t] \begin{center} 
\epsfig{figure=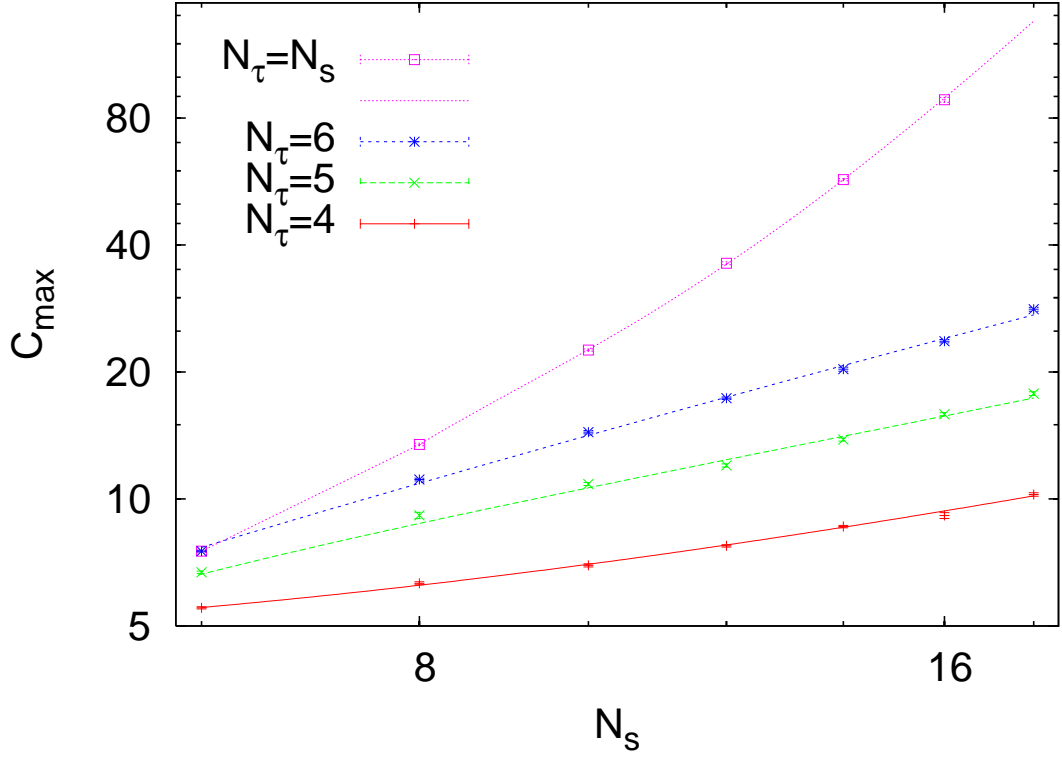,width=\columnwidth} \vspace{-1mm}
\caption{Maxima of the specific heat. \label{fig_Cmax} }
\end{center} \vspace{-3mm} \end{figure}

\begin{figure}[-t] \begin{center} 
\epsfig{figure=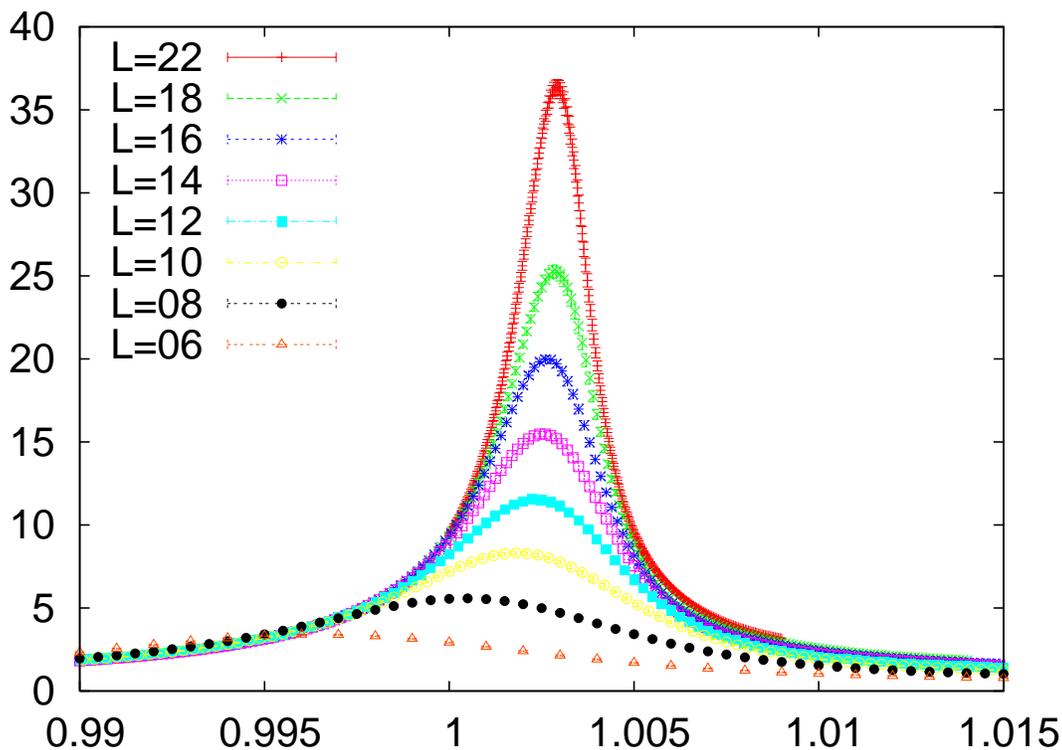,width=\columnwidth} \vspace{-1mm}
\caption{Polyakov loop susceptibilities for $N_{\tau}=4$. 
\label{fig_CP04} }
\end{center} \vspace{-3mm} \end{figure}

The lattice size dependence of the maxima of the specific heat is
shown in Fig.~\ref{fig_Cmax}. For our symmetric lattices a fit to 
the first-order transition form
\begin{equation} \label{Cmax1}
   C_{\max}(N_s)/(6N)=c_0+a_1/N+a_2/N^2  
\end{equation} 
yields the estimate with $c_0\approx 0.00020$. Statistical error bars
of our estimates can be found in \cite{BeBa06}. Our $c_0$ value is 
10\% higher than the one reported in Rev.~\cite{ArBu03}, where lattices 
up to size $18^3$ were used for this fit. 

In contrast to symmetric lattices we find for $N_{\tau}=4$, 5 and~6 from 
Eqs.~(\ref{Cbeta}), (\ref{chi}), (\ref{chi_beta}) and~(\ref{sffss})
\begin{equation} \label{exponents}
  \frac{\alpha}{\nu}\approx 1\,,\qquad 
  \frac{\gamma}{\nu}\approx 2\,,\qquad
  \frac{1-\beta}{\nu}\approx 1.5~~~~{\rm and}~~~~
  2-\eta \approx 2\ .
\end{equation} 
The hyperscaling relation $2-\alpha=d\nu$ with $d=3$ implies then
$\alpha\approx \nu \approx 0.5$. For the Polyakov loop susceptibility 
and $N_{\tau}=4$ Fig.~\ref{fig_CP04} shows our canonically reweighted 
data, implying $\gamma/\nu=2$ and, hence, $\gamma\approx 1$. For the 
reweighting the logarithmic coding from chapter~5.1.5 of \cite{BBook} 
was used.  From the derivative of $\langle|P|\rangle$ we get similarly 
$\beta\approx 0.25$, which is consistent with the scaling relation 
$\alpha+2\beta+\gamma=2$. While from the structure factor we find 
$2-\eta$ consistently with the scaling relation $\gamma/\nu=2-\eta$.

The exponents listed are the 3d Gaussian values. In view of statistical 
errors and expected systematic errors due to our limited lattice sizes, 
close-by values are as well possible. The problem with the Gaussian 
values is that the Gaussian renormalization group fixed point in 3d 
has two relevant operators \cite{Gaussian}. So it is unstable and 
one does not understand why the effective spin system should care 
to converge into it \cite{SvYa82}. Therefore, the interesting scenario 
of a new non-trivial fixed point with exponents accidentally close to 
3d Gaussian arises. An illustration, which is consistent with the 
data, are the values $\nu=0.482$, $\alpha=0.554$, $\gamma=0.94$, 
$\beta=0.253$, $\eta=0.05$. Of course, we cannot entirely exclude 
that finite lattices mislead us and that really large systems turn 
either towards a first order transition or the 3d XY fixed point.

\begin{figure}[-t] \begin{center} 
\epsfig{figure=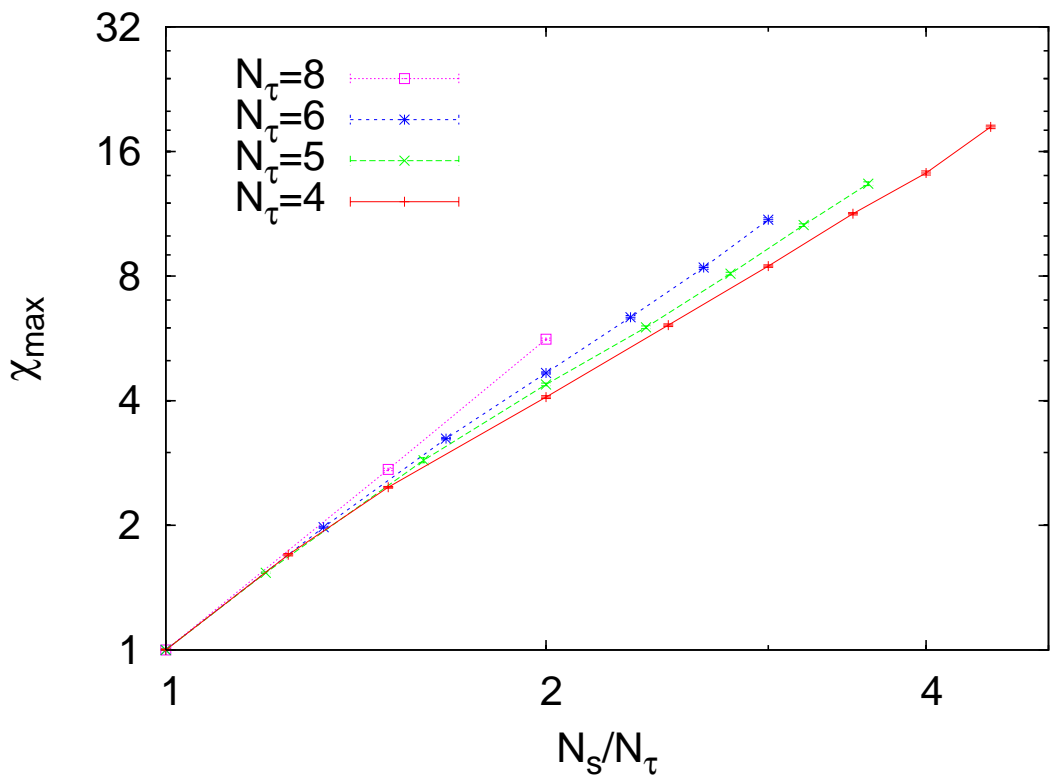,width=\columnwidth} \vspace{-1mm}
\caption{Rescaled maxima of Polyakov loops susceptibilities. 
\label{fig_CPnorm} }
\end{center} \vspace{-3mm} \end{figure}

U(1) LGT in the $N_t\,N_s^3,\,N_s>N_{\tau}$ geometry has previously
only been studied by Vettorazzo and de Forcrand \cite{VeFo04}. While
their result for $N_{\tau}=4$ are consistent with a second order
transition, they claim first order transitions for $N_{\tau}=6$ 
and~8. However, their $N_{\tau}=6$ and~8 data on very large lattices 
may simply exhibit the critical slowing down, which is typical for 
second order transitions \cite{BeBa06}. For $N_{\tau}=4,$ 5 and~6 our 
data support, independently of $N_{\tau}$, the same critical exponent.
We illustrate this here by rescaling the maxima of our Polyakov loop 
susceptibilities with a common factor, so that they become equal to~1 
on symmetric lattices. On a log-log scale the results are then plotted 
in Fig.~\ref{fig_CPnorm} against $N_s/N_{\tau}$. The behavior is 
certainly consistent with assuming a common critical exponent for 
all of them (parallel lines are expected for large $N_s/N_{\tau}$). 
The figure includes also some preliminary $N_{\tau}=8$ data, which 
so far blend in. 

\section{Summary and Conclusions}

The litmus test for identifying a second order phase transitions
is that one is able to calculate its critical exponents unambiguously. 
With MC calculations and FSS methods this finds its limitations 
through the lattice sizes, which fit into the computer and can be 
simulated in a reasonable time. Our lattice sizes are not small on 
the scale of typical numerical work on U(1) LGT, for instance the 
lattices used for the $c_0$ estimate of \cite{ArBu03}. However, in 
view of the fact that our data do not support the generally expected 
scenarios, it would be desirable to extent the present analysis to 
lattices up to, say, size $32^4$. Using supercomputers this appears 
feasible. With mass spectrum methods \cite{BePa84}, see \cite{recent} 
for recent work, one may investigate the scaling behavior of the model 
from a different angle. Finally, renormalization group theory could 
contribute to clarifying the issues raised by our data.

\acknowledgments
BAB thanks Thomas Neuhaus for communicating and clarifying details
of the Wuppertal data. We thank Philippe de Forcrand for e-mail 
exchanges and useful discussions, and Urs Heller for checking on 
one of our action distributions with his own code. This work was 
in part supported by the US Department of Energy under contract 
DE-FA02-97ER41022.

\end{document}